\documentclass[conference]{IEEEtran}
\IEEEoverridecommandlockouts
\usepackage{cite}
\usepackage{amsmath,amssymb,amsfonts}
\usepackage{algorithmic}
\usepackage{graphicx}
\usepackage{textcomp}
\usepackage{xcolor}
\usepackage{url}
\usepackage{bm}

\def\BibTeX{{\rm B\kern-.05em{\sc i\kern-.025em b}\kern-.08em
    T\kern-.1667em\lower.7ex\hbox{E}\kern-.125emX}}
\begin{document}

\title{Analysis of Arctic Buoy Dynamics using the Discrete Fourier Transform and Principal Component Analysis\\
}

\author{\IEEEauthorblockN{James H. Hepworth}
\IEEEauthorblockA{\textit{	Department of Mechanical Engineering} \\
\textit{University of Cape Town}\\
Cape Town, South Africa \\
james.hepworth@uct.ac.za}
\and
\IEEEauthorblockN{Amit Kumar Mishra}
\IEEEauthorblockA{\textit{Department of Electrical Engineering} \\
\textit{University of Cape Town}\\
Cape Town, South Africa\\
akmishra@ieee.org}
}

\maketitle

\begin{abstract}
Sea-Ice drift affects various global processes including the air-sea-ice energy system, numerical ocean modelling, and maritime activity in the polar regions. Drift has been investigated via various technologies ranging from satellite based systems to ship or ice-borne processes. This paper analyses the dynamics of sea-drift in the Arctic over 2019-2021 by Fourier Analysis and Principal Component Analysis of displacement data generated from the drift tracks of Ice-Tethered Profilers. We show that the frequency characteristics of drift support the notion that it is a function of both slowly varying processes, and higher frequency, random, forcing. In addition, we show that displacement data features high correlation between deployment locations and, consequently, suggest that there is scope for the optimisation of profiler deployment locations and for the reduction in number of instruments required to capture the displacement characteristics of drift.
\end{abstract}

\begin{IEEEkeywords}
Ice-Tethered Profilers, Sea-Ice Drift, Fourier Analysis, Principal Component Analysis
\end{IEEEkeywords}

\section{Introduction}\label{intro}

Sea-ice drift is a phenomenon of interest as a scientific parameter due to its role in the air-sea-ice energy systems at the poles, its role in high resolution regional ocean models, and as it relates to the operation of marine vessels in ice-covered waters \cite{Lund2018-ys}. Since the 1980s sea-ice drift has been understood to be a combination of seasonal (or longer time scale) effects which are responsible for slowing changing, mean motion, and rapidly changing, higher frequency, random components \cite{Thorndike1986-hg, Rampal2009-ae}.  Low-frequency effects are the response to the mean effect of phenomena such as ocean currents \cite{Rampal2009-ae}, wind \cite{Tandon2018-vl}, and tidal forcing \cite{Heil2008-om}, however the high-frequency artifacts present in sea-ice drift velocities in the Arctic have been shown to be multi-fractal and more complex than can be explained purely be forcing due to the ocean and atmosphere \cite{Rampal2009-ae, Rampal2019-lz,  Weiss2009-ll}.

Several remote sensing technologies have been employed to assess sea-ice drift: These include various satellite-based approaches such visual and Infra-Red spectrum imaging \cite{Emery1986-dp}; Synthetic Aperture Radar \cite{Curlander1985-rx, Holt1992-fw, Bouillon2015-bl} and Passive Microwave imaging \cite{Kwok2010-vt, Sumata2014-ho}; whilst ship borne marine radar \cite{Lund2018-ys} has recently been used for the analysis of sub-mesoscale drift at high resolutions. In addition to these, the use of in-situ data sets generated by ice-tethered instruments which drift with the ice and act as Lagrangian trackers, provide useful data from which drift can be directly investigated.

Ice-Tethered Profilers (ITPs) are in-situ, ocean monitoring instruments designed to sample temperature and salinity of the ocean at a range of depths up to 800 m below the surface of the ocean via a sensor unit able to travel a tether suspended from an ice-floe mounted surface unit \cite{Toole2011Ice, ITPs}. The surface unit collects the Global Positioning System (GPS) locations of top of the tether and thus, in addition to temperature and salinity depth profiles, sea-ice drift can be tracked against time from the Eulerian viewpoint of the GPS. Several of these instruments have been successfully deployed in the Arctic each year since the mid-2000s resulting in over 100 unique data sets being generated: Several are ongoing whilst numerous missions are complete. In this paper, the terms, ``buoys'' and ``ITPs'' are used synonymously.

In this paper we investigate the latest available data sets from completed ITP missions in the Arctic Ocean for instrument deployments spanning the period from October 2019 to August 2020 and again from late October 2021 until the end of that year.  The motion characteristics of ice drift over this period are determined using two methods: Fourier analysis of discrete buoy displacement data sets and Principal Component Analysis (PCA) of the same data.

Here we show, first, that the drift of buoys deployed across the Arctic show similar primary frequency components which do not depend on the actual location of the instrument and the corresponding geometry of the drift track. Second, our results confirm the notion of sea-ice drift being a function of two parts: slow, low frequency drivers governing mean motion, and high frequency additions responsible for the random motion of ice at short time scales.  Finally, we show a high degree of correlation within data from concurrently deployed buoys and, hence, make the argument that the number of deployed buoys can be reduced if specific experimental goals can be met using analysis of displacement derived from the drift profiles, such as experiments which investigate the rates of surface drift for various regions.

The remainder of this paper is organised as follows. First, in in Section \ref{methodology}, the methodological approach used to perform the investigation is presented, including the source of data and its pre-processing.  The particular drift profiles in this analysis are then shown in Section \ref{Drift} before the results of Fourier Analysis and PCA are shown in Sections \ref{FA_analysis} and \ref{PCA_analysis}, respectively.  Finally, the contributions made by this paper are discussed and concluded in Section \ref{Disc_Conc}.

\section{Methodology}\label{methodology}

\subsection{Source of data sets}\label{data_source}

In this paper, drift tracks and sequential displacement data sets of ITPs were developed from the raw GPS position data of ITPs deployed in the Arctic Ocean between 2019 and 2021.  The Ice-Tethered Profiler data were collected and made available by the Ice-Tethered Profiler Program  based at the Woods Hole Oceanographic Institution (https://www2.whoi.edu/site/itp/) \cite{Toole2011Ice, Krishfield2008Automated}.  The full data sets for the years of analysis in this paper as well as those for all other completed missions are available on request.

\subsection{Pre-processing of data}\label{pre_proc}

For the purpose of this analysis, only completed ITP mission data sets from the 2019-2021 period are used. GPS locations are sampled from the buoys at varying frequencies with different buoys using a variety of nominal sample frequencies ranging from every 30 minutes to every 6 hours. Additionally, all raw data sets show gaps where GPS locations were unable to be determined for several sample periods. Both Fourier analysis and time-domain PCA require constant sampling rates, whilst PCA also requires parallel samples to be taken at the \textit{same} instant in time. Thus, the data required pre-processing to ensure that position measurements were rounded to a equatable time instances and to ensure that each data set contained entries at equal sample rates. Accordingly, the raw data sets were pre-processed as shown by the flowchart in Figure \ref{fig1} below.

\begin{figure}[!bp]
  \centering
  \includegraphics[width=1\linewidth]{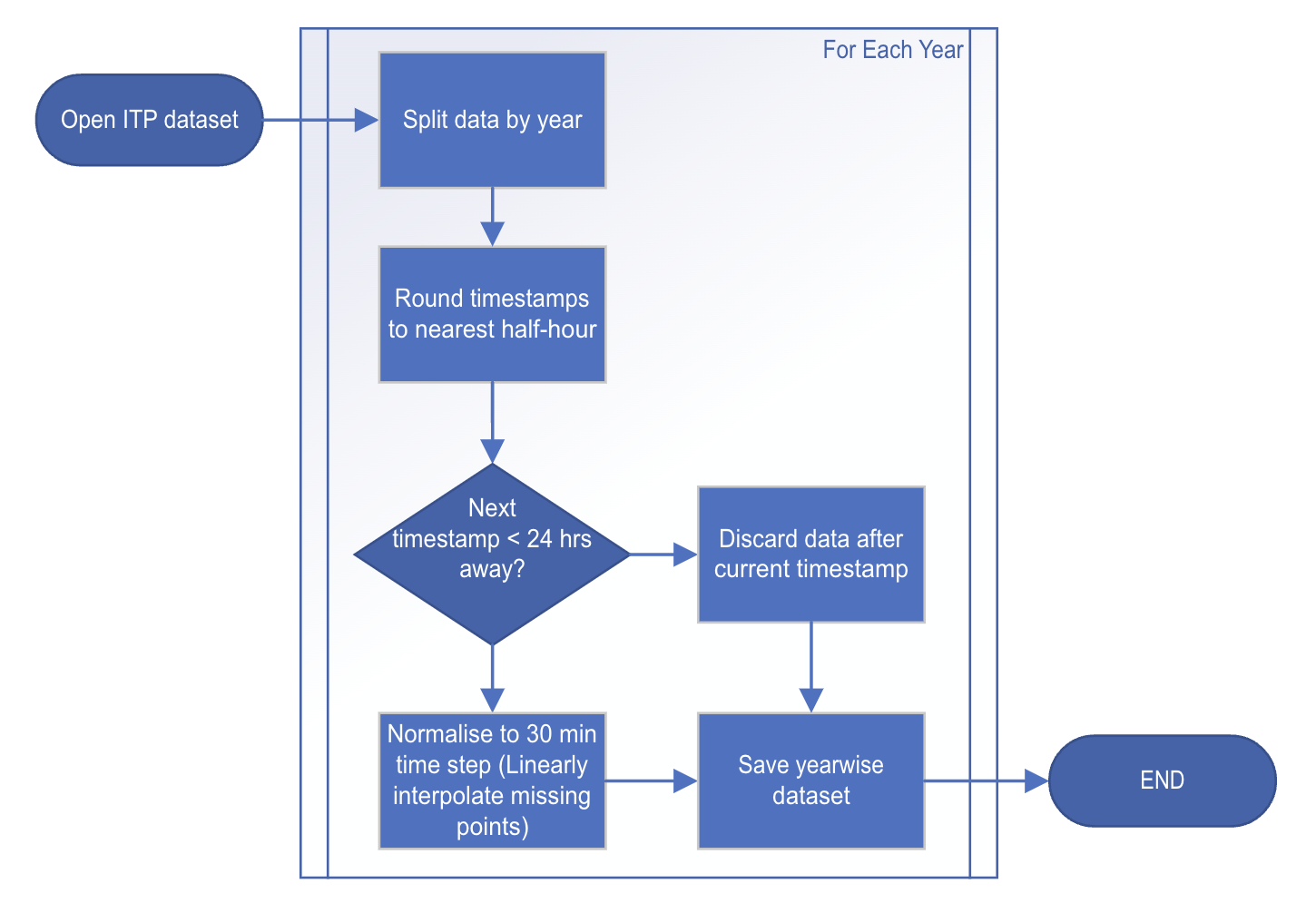}

  \caption{
    Pre-processing algorithm applied to the raw locations data set from each completed ITP mission.
  }
  \label{fig1}
\end{figure}

As Figure~\ref{fig1} indicates, data sets were abandoned after a time gap of more than 24 hours was detected. From the collection of new data sets produced, each containing data points separated by 30 minutes on the daily half-hour marks (UTC) the analyses described in Sections \ref{FA_analysis} and \ref{PCA_analysis} below were conducted.

\subsection{Fourier Analysis}\label{Fourier}

The Fourier transform is used, in general, to decompose, continuous, cyclic signals in the time-domain into a sum of sinusoids whose amplitudes and can be represented in the frequency domain. This allows investigation into the effects of processes occurring at particular frequencies on the motion as a whole to be evaluated. However, the data used in this investigation are discrete and finite rather than continuous and cyclic, with buoy locations starting in one position and ending at another in the data set. Therefore, to account for the discontinuity in the data sets between these start and end locations, a Hanning window, as shown in Equation \ref{eq1}, below, was used in conjunction with the Discrete Fourier Transform (DFT).

\begin{equation}
\label{eq1}
w(i) = 0.5 - 0.5 \cos \left( \frac{2\pi i}{N-1} \right), \quad 0 \le i \le N- 1
\end{equation}

Here, $w(i)$ is the magnitude of the window output, \textit{i} is the instantaneous sample number of the window, and \textit{N} is the length of the window output data set, necessarily of the same length as the displacement data set under evaluation. The DFT was  applied to the data according to Equation \ref{eq2}, below \cite{DFT}:

\begin{equation}
\label{eq2}
|X_k| =  \frac{1}{N} \sum_{n=0}^{N -1} x_n e^{ -i 2\pi k n w(i)/N }, \quad k = 0, \ldots, N -1.
\end{equation}

Here, \textit{N} is, again, the number of entries in the data set, \textit{n} is the instantaneous sample, \textit{k} is the instantaneous frequency, \textit{x\textsubscript{n}} is the sinusoidal value of the instantaneous sample, and $ |X_k|$ is the amplitude of the signal. 

\subsection{Principal Component Analysis}\label{PCA}

PCA is a data transformation technique used to determine the optimal bases of data sets such that maximal variance in the data is captured using the fewest basis vectors. Individual ITP data sets (time series) were first collected into a matrix, $\bm{X}$, with each column representing each sensor/data set, and then centred about their mean values, $\bm{\mu}$, and standardised by their standard deviations, $\bm{\sigma}$, to eliminate the effect of possible wide variation of magnitude in the original data. This produces the matrix, $\bm{Z}$. The covariance matrix of $\bm{Z}$, the correlation matrix, was then calculated, before the matrix of eigenvectors, $\bm{P}$, were determined according to Equation \ref{eq3}, below:

\begin{equation}
\label{eq3}
\bm{corr} = \bm{Z^TZ} = \bm{PDP^{ -1}}
\end{equation}

Here, $\bm{D}$ is a diagonal matrix of the eigenvalues corresponding to the eigenvectors $\bm{P}$. Once the vectors of $\bm{P}$ are sorted according to the descending order of the eigenvalues in $\bm{D}$ to form $\bm{P'}$,  the centred representation of $\bm{X}$ according to the new basis vectors given by the eigenvectors of the correlation matrix of $\bm{Z}$ is determined, $\bm{Z'}$, according to Equation \ref{eq4}, below:

\begin{equation}
\label{eq4}
\bm{Z'}=\bm{PZP'}
\end{equation}

The reconstruction of data, $\bm{X'}$, according to the original bases can be accomplished according to Equation \ref{eq5}, below:

\begin{equation}
\label{eq5}
\bm{X'}=\bm{Z'P'^{T}}\bm{\sigma} + \bm{\mu}
\end{equation}

\section{Drift of ITPs in the Arctic}\label{Drift}

\begin{figure}[!bp]
  \centering
  \includegraphics[width=1\linewidth]{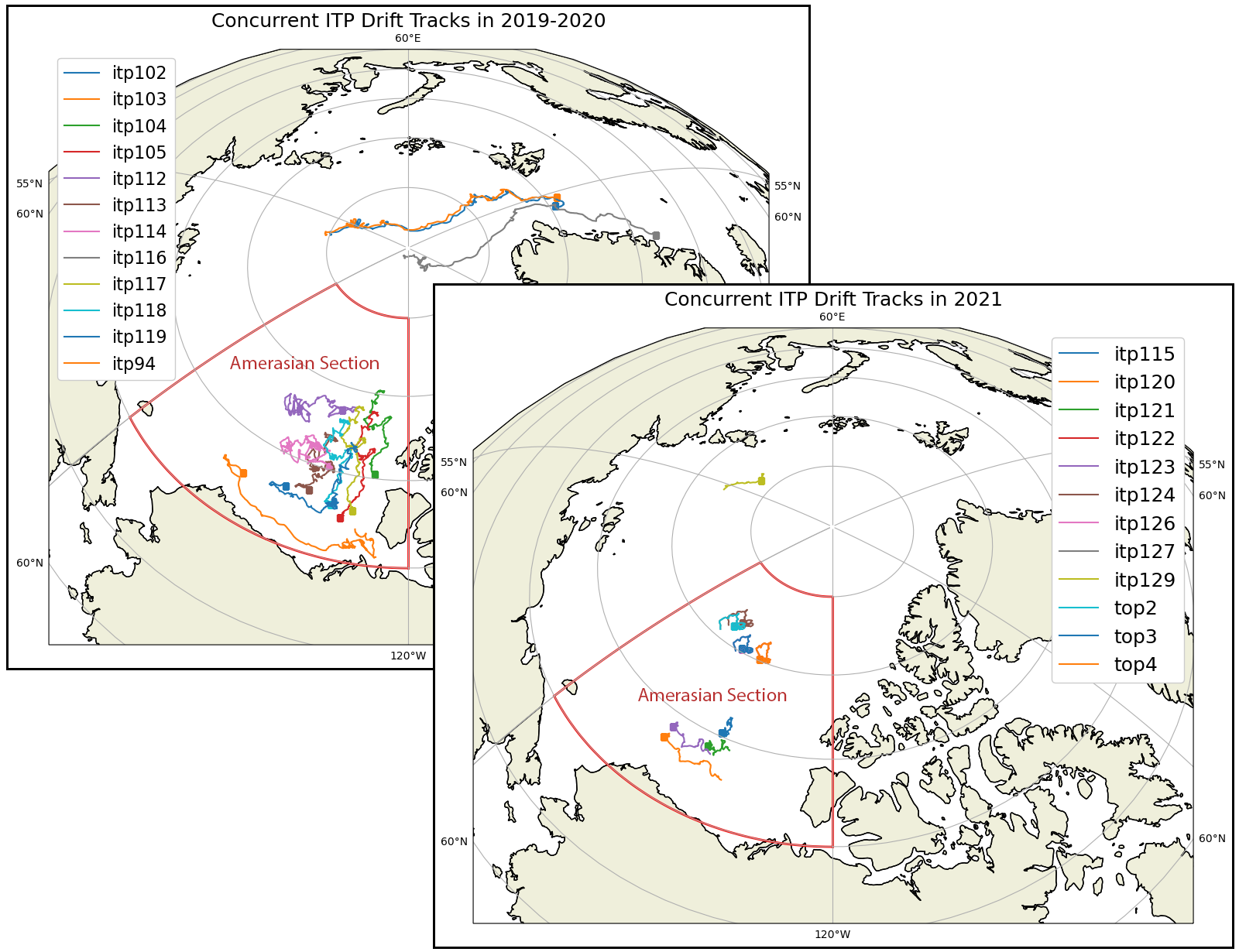}

  \caption{
    Drift tracks of concurrently deployed ITPs in the Arctic in years 2019-2020 (top) and 2021 (bottom). The end locations of the instruments are shown with a square marker.
    }
  \label{fig2}
\end{figure}

The drift tracks shown in Figure~\ref{fig2}, below, are those of concurrently deployed ITPs in the Arctic in 2019-2020, and 2021. From these figures it is evident, that, broadly speaking, drift behaviour is dependent on location with buoys deployed in the Eurasian Basin showing more linear translation southward (e.g. ITPs 102 and 116 in 2020), and buoys deployed in the Amerasian Basin showing a less linear drift (e.g. top2 and top4 in 2021). From these drift tracks, the half-hourly displacement (geodesic distance) was calculated between consecutive locations to produce a displacement data set. 

Due to the significant differences in apparent motion for buoys in the Amerasian Basin from those in the Eurasian Basin, the displacement data sets were then split into two data sets: The first, termed the Amerasian section in this paper, consisting of buoys deployed in the window bounded by 180-120 ºW and 65-85 ºN, and the second, termed the Eurasian section, for those not deployed in that window. These location specific data sets of ITP displacements every half-hour were then used in the analyses of Section~\ref{FA_analysis} and Section~\ref{PCA_analysis}.

\section{Fourier Analysis}\label{FA_analysis}

\begin{figure}[!tbp]
  \centering
  \includegraphics[width=1\linewidth]{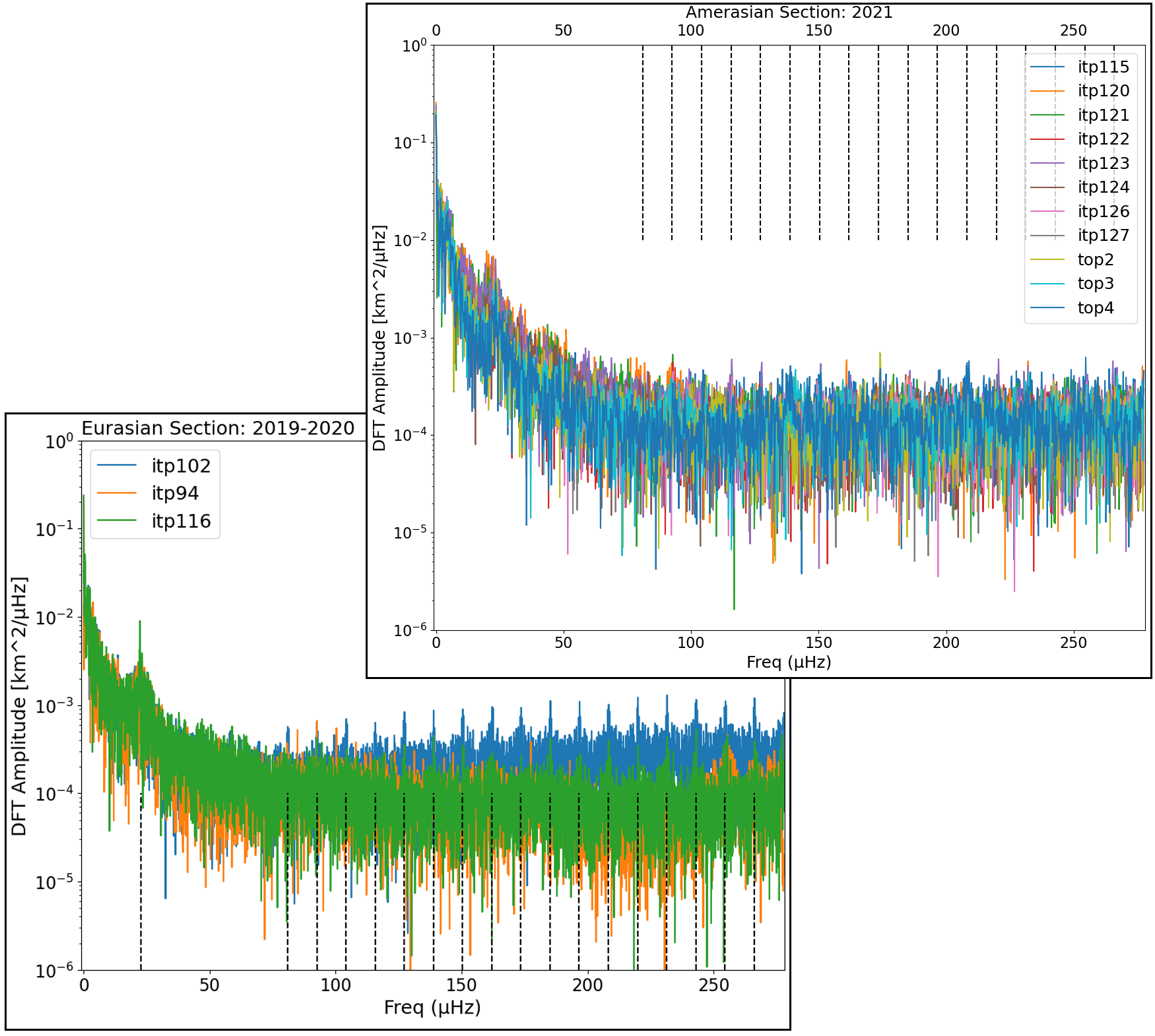}

  \caption{
    Normalised DFTs of the ITP displacements for the Amerasian section in 2021 (top), and the Eurasian section in 2019-2020 (bottom).
    }
  \label{fig3}
\end{figure}

The DFT described in Section~\ref{Fourier}, above, was applied to the zonal and meridional displacement data sets, via a Hanning window, and normalised by the number of samples to produce the frequency responses shown in Figure \ref{fig3}, above.  These plots, whilst only showing the responses for buoys in the Eurasian section in 2019-2020 and the Amerasian section in 2021, are representative of those from the corresponding sections in the years not shown as well.

The description of sea-ice drift in the Arctic as the sum of slow, seasonal (or longer), processes and high frequency fluctuations is supported by the results of Figure~\ref{fig3}. Here, it is clear from the low frequency peaks, predominantly between 10\textsuperscript{-1} and 10\textsuperscript{0} km\textsuperscript{2}/$\mu$Hz, that displacement is primarily driven by slow, DC, processes. However it can also be seen that a secondary dominant component of the order of 10\textsuperscript{-2} km\textsuperscript{2}/$\mu$Hz is evident at {$\sim$}22.6 $\mu$Hz (or every 12.3 hours) which corresponds well with the tidal frequency \cite{Tides}. In addition to these primary features in the data, of interest are the peaks in almost all ITP data sets of {$\sim$}10\textsuperscript{-3} km\textsuperscript{2}/$\mu$Hz which begin at {$\sim$}81 $\mu$Hz (or 3.4 hours) and repeat with increasing frequency with every {$\sim$}11.6 $\mu$Hz above this threshold until the Nyquist frequency for the data is reached at {$\sim$}277 $\mu$Hz (or every 1 hour).

\section{Principal Component Analysis}\label{PCA_analysis}

PCA was then applied to the same displacement drift data sets as used in the Fourier analysis above, and applied separately to the Eurasian and Amerasian sections for each year as before.  Presented in Table \ref{tab1}, overleaf, are the percentages of the cumulative variance captured by the first two principal components (PCs) for the data assessed. 

\begin{table}[!tp]
  \centering
  \caption{
    Cumulative variance of ITP displacement  captured by the first two PCs
  }
  \label{tab1}

  \begin{tabular}{p{\dimexpr 0.333\linewidth-2\tabcolsep}p{\dimexpr 0.333\linewidth-2\tabcolsep}p{\dimexpr 0.333\linewidth-2\tabcolsep}}
    \hline
    \textbf{Year} & \textbf{Eurasian Section} & \textbf{Amerasian Section} \\
    \hline
    2019-2020 & 95.3\% & 74.5\% \\
    2021 & 100\% & 83.9\% \\
    \hline
  \end{tabular}
\end{table}

Reconstruction of the displacement data sets using only the first two PCs was then performed. Figure~\ref{fig4}, overleaf, shows the similarity of the reconstructed displacement signals with the original data. Here it can be seen that the plot of ITP 103 has been reconstructed more effectively than that of ITP 104. Quantitatively, the RMS error in reconstruction of ITP 103 is 0.030 km compared to 0.078 km for ITP 104. This is due to the magnitudes of ITP 103's motion being greater than 104's and thus contributing more to the total variance of the original data set. Therefore, its motion is more likely to be captured by the first two PCs which capture the dominant sources of variation. 

\begin{figure}[!b]
  \centering
  \includegraphics[width=1\linewidth]{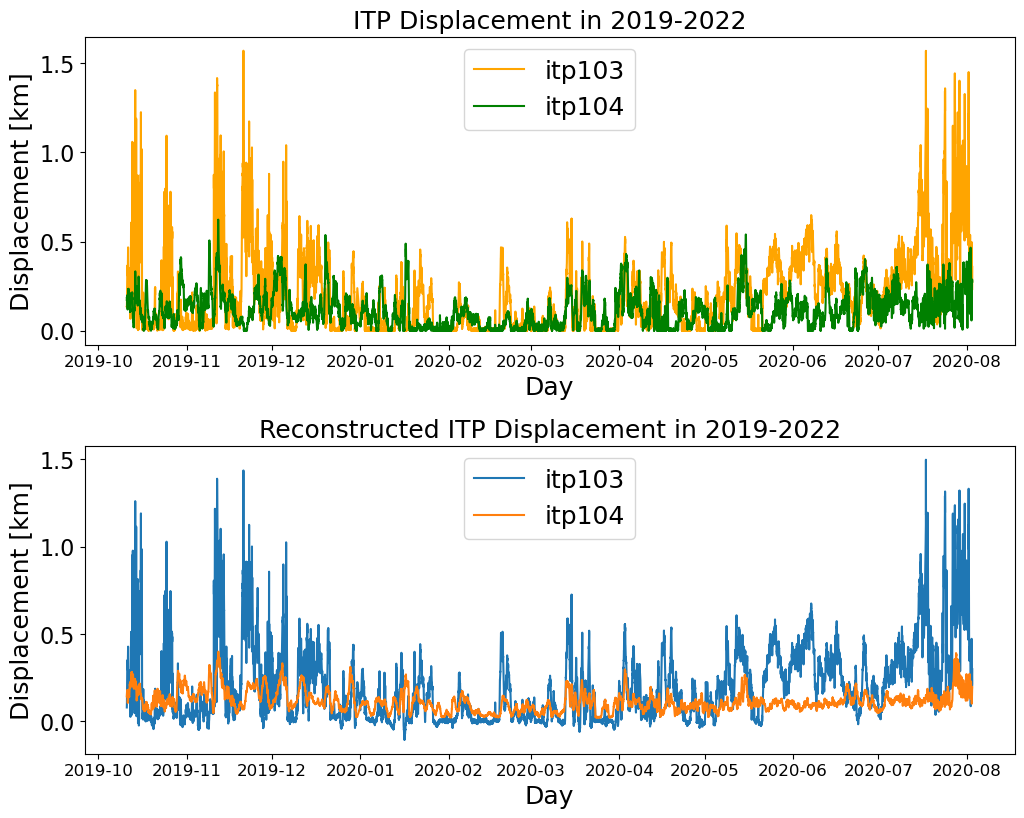}

  \caption{
    Comparison between selected original data plots of two ITPs in 2019-2020 (top) with the reconstructed plots formed using two PCs.
  }
  \label{fig4}
\end{figure}

\begin{figure}[!h]
  \centering
  \includegraphics[width=1\linewidth]{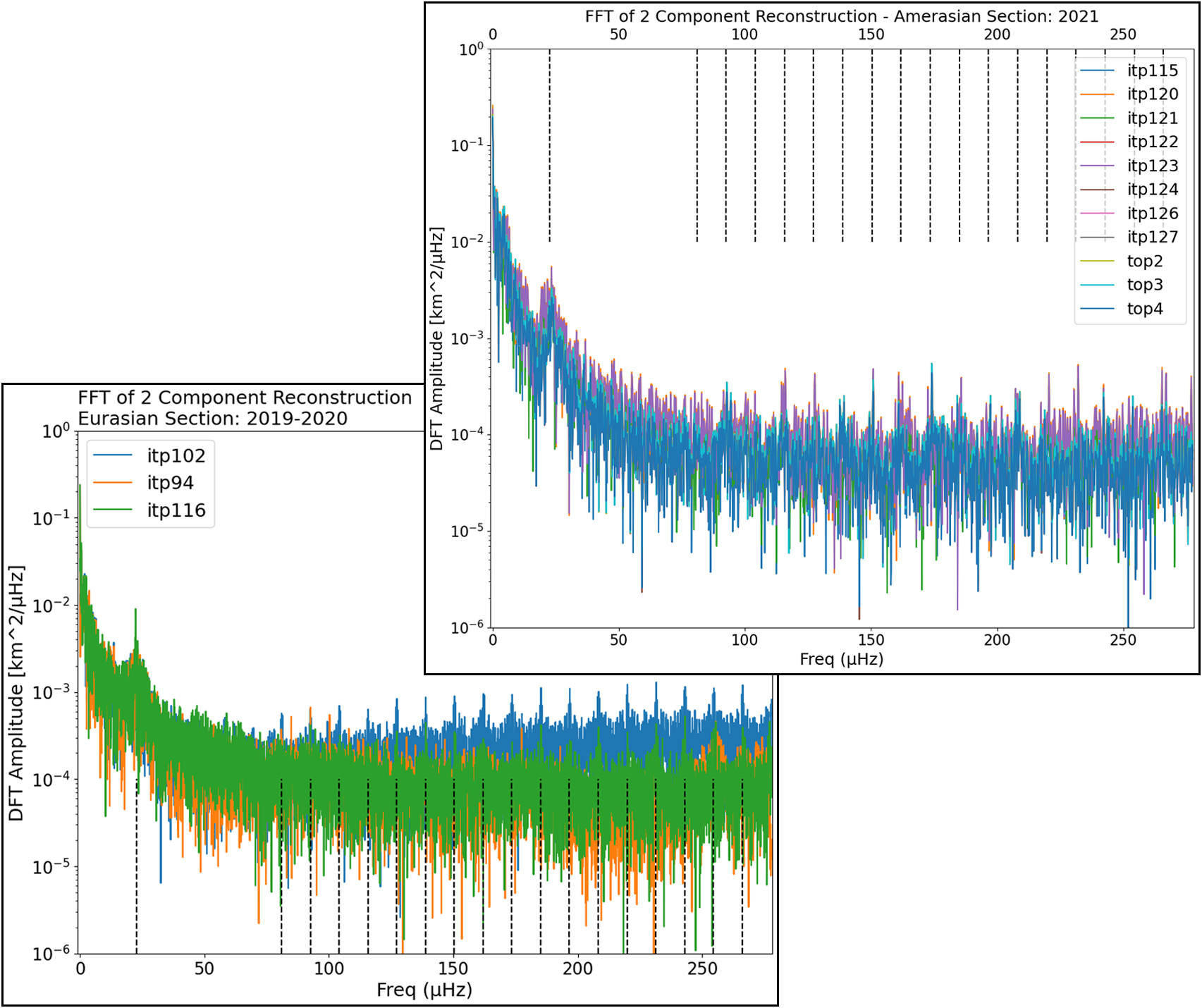}

  \caption{
    DFT results from the displacement data reconstructed from the first two PCs for the 2019-2020 Eurasian ITPs (bottom) and 2021 Amerasian ITPs (top).
  }
  \label{fig5}
\end{figure}

The DFT was also applied to these reconstructed data sets in the same manner shown in Section~\ref{FA_analysis}. The results of these transforms are shown in Figure~\ref{fig5}, opposite. It's similarity with Figure~\ref{fig3} suggests that the frequency characteristics of the displacement data sets consisting of (in these cases) up to 10 ITPs can be approximated with far fewer, optimal, drifters. I.e. those that would have drifted according to a pattern sufficient to generate the displacements given by the first two PCs. 

Further, during the PCA analysis, the coefficients of correlation, \textit{r}, between distinct ITPs were resultant. ITPs in close vicinity of each other show high degrees of correlation in their drift motion whilst those separated by great distances are not correlated. Table~\ref{tab2}, opposite,  shows the \textit{r} values between the displacements of selected ITPs. ITPs 102 and 94 (the blue and orange tracks at the top of Figure~\ref{fig2}) which overlap and follow very similar trajectories in 2019-2020 are highly correlated. Conversely, ITPs 102 and 94 are highly uncorrelated with ITP 112 (purple in Figure~\ref{fig2}). Similar results are evident for 2021 as well.

\begin{table}[!t]
  \centering
  \caption{
    Correlation coefficients for selected ITPs in 2019-2020
  }
  \label{tab2}

  \begin{tabular}{p{\dimexpr 0.250\linewidth-2\tabcolsep}p{\dimexpr 0.250\linewidth-2\tabcolsep}p{\dimexpr 0.250\linewidth-2\tabcolsep}p{\dimexpr 0.250\linewidth-2\tabcolsep}}
    \hline
    \textbf{ITP} & \textbf{102} & \textbf{94} & \textbf{112} \\
    \hline
    102 & 1 & 0.841 & 0.039 \\
    94 & 0.841 & 1 & 0.052 \\
    112 & 0.039 & 0.052 & 1 \\
    \hline
  \end{tabular}
\end{table}

\section{Discussion and Conclusion}\label{Disc_Conc}

In this paper we have described the process followed to investigate the drift of ITPs deployed in the Arctic over the years 2019-2021 via Fourier analysis and PCA of the displacement data sets derived from the time series of GPS locations available for each completed ITP mission in that period. The drift plots shown in Section III highlight the fact that drift of sea-ice in the Arctic is highly variable between significantly separated buoys but can show high similarity in profile if the points of observation are close together.

Section IV analysed the frequency components of the displacement data using the DFT. Here, it was shown that, despite the vastly different different drift profiles evident across the Arctic sections, similar processes drive the displacement of sea-ice: The low-frequency primary component of the signals at $\sim$0 Hz can be attributed to the average of long-term features in the environment such as ocean currents and prevailing winds [2, 19], whilst the second dominant peak at $\sim$22.6 $\mu$Hz aligns with the tidal frequency. Together these support the notion that sea-ice drift is primarily driven by slow processes. Of interest are the repeating peaks in the DFT outputs of all displacement data every 11.6 $\mu$Hz increasing from $\sim$81 $\mu$Hz. Physical drivers of these features have not been determined and are the ongoing subject of further investigation. The nature of the DFT results here have implications for ice-tethered sensor design and their sampling strategy when drift analysis is the primary experimental goal; most notably that they should be optimised for low frequencies due to the fact that the low frequency features in the DFT response are 1-3 orders of magnitude greater in amplitude than those which repeat with increasing frequency. Sensors can, therefore, be sampled at low frequencies with low-power states being entered between sampling to reduce their power demands and increase their battery run time whilst still capturing the dominant drivers of drift. In addition, the use of low-pass filtering can be employed to reduce the influence of noise on the data obtained. Further, due to the similarities in the DFT results across all regions of the Arctic, one can have confidence that a drift sensor design and sampling strategy developed for deployment in one region will also be suitable for any other region.

Section V applied PCA to the data sets and showed, empirically, that high correlation in drift dynamics are evident between sensors deployed in near vicinity to each other. This was shown by the correlation coefficients of example buoys approaching 1 in these cases. This section also showed that the displacement data from concurrently deployed buoys can be reconstructed with high accuracy from only two principal components. This implies that fewer buoys can be deployed for experiments which aim to analyse the displacement of sea-ice drift from in-situ sensors (for example \cite{Timmermans2008-il}). This has potential implications to both reduce the cost of experiments through the use of fewer buoys if their deployment locations are optimally chosen, and to and increase the efficiency of experimental deployments from time and resource constraint perspectives. These results help to present the case for the development of a methodology to optimise the choice of deployment locations for ice-tethered sensors measuring sea-ice drift. This subject is the immediate focus of the authors attention.

Further, for a given experimental monetary budget, the fewer the number of buoys that are required to achieve a particular experimental goal, the greater the proportion of the budget that can be allocated to the engineering requirements of the project. I.e higher quality sensors or materials could be used to increase the confidence or accuracy of data generated, or the lifespan of the system. Several studies highlight that despite the wealth of existing historical data from the Arctic, both from remote sensing and in-situ methods, there is a need for increased measurements to provide the required data to better understand the rapidly changing environment of the Arctic \cite{Timmermans2020-iz,Timmermans2018-yw,Nguyen_undated-gq}. Consequently, optimising the deployment of buoys such that high quality data may be timeously obtained across a wide area, is a pressing need in order to provide these data to the physical and numerical oceanography scientific communities.

These factors, discussed above, also apply to the study of the Southern Ocean (SO), where they are, arguably, of even greater concern. The SO has a significant effect on the major ecological, climate and environmental systems of the Earth, however, the SO is simultaneously one of the most under-sampled and least observed environments on Earth, leading to large uncertainties with and wide variations
between the outputs of various climate related products \cite{Smith2019-xw}. This under-sampling is primarily due to the extreme remoteness and harshness of the environment making it difficult and expensive to access and to deploy in-situ sensors \cite{Newman2019-aq, Newman2016-tz, Barbariol2019-wz}. Our own experience has shown that ice-tethered sensors which are routinely deployed in the Arctic do not survive the conditions of the Antarctic well. Thus, ourselves and other researchers in this field have resorted to the development of specifically designed buoys to measure drift and other features in the Southern Ocean \cite{Jacobson2021-vp, Kohout2015-ae}. In these cases where deployment is difficult and costly, and turnkey sensor solutions are not suitable, optimisation of sensor deployment and project budgets is of great potential benefit. Accordingly, further analysis is underway to determine the applicability of the results shown in this paper, and any resultant deployment optimisation methodology developed, to regions of the Antarctic ocean where the characterisation of the free-drift of sea-ice is primitive and a subject of active research \cite{De_Vos2022-qe,Womack2022-bb}.

\section{Acknowledgement}
This work has been supported with funding from Sentech Soc Ltd, South Africa.

\bibliographystyle{IEEEtran}
\bibliography{main}
\end{document}